%% file: main.tex
\documentclass[10pt,twocolumn,aps,prb,superscriptaddress,footinbib]{revtex4-2}
\usepackage{amsfonts}
\usepackage{amsmath}
\usepackage{amssymb}
\usepackage{color}
\usepackage{graphicx}
\usepackage{bm}
\usepackage{xcolor}

\usepackage{soul}
\usepackage{cancel}
\usepackage[colorlinks,bookmarks=false,allcolors=blue,hyperfootnotes=true]{hyperref}
\usepackage{float}				
\usepackage{multirow}

\begin{document}

\title{Multiple Quasiparticle Bound States in a Trap Created by a Local
Superconducting Gap Variation}

\date{\today}

 \author{Romy Morin}
 \affiliation{Univ.~Grenoble Alpes, CEA, Grenoble INP, IRIG, PHELIQS, 38000 Grenoble,
France}

 \author{Denis M.~Basko}
 \affiliation{Univ.~Grenoble Alpes, CNRS, LPMMC, 38000 Grenoble,
France}

  \author{Manuel Houzet}
  \affiliation{Univ.~Grenoble Alpes, CEA, Grenoble INP, IRIG, PHELIQS, 38000 Grenoble,
France}

 \author{Julia S.~Meyer}
 \affiliation{Univ.~Grenoble Alpes, CEA, Grenoble INP, IRIG, PHELIQS, 38000 Grenoble,
France}

\begin{abstract}
At low temperature, the concentration of quasiparticles observed in superconducting circuits far exceeds the predictions
of microscopic BCS theory at equilibrium. As a source of dissipation, these excess quasiparticles degrade the
performance of various devices. Therefore, understanding their dynamics, especially their recombination into Cooper
pairs, is an active topic of current research. In disordered superconductors, spatial fluctuations in the superconducting
gap can trap quasiparticles and modify their eigenspectrum. Since this spectrum plays a key role in quasiparticle
dynamics, it must be carefully investigated. To this end, we introduce a toy model of a single trap. Specifically, we
consider a shallow disk-shaped gap variation in a clean superconductor. Using a semiclassical approximation,
we demonstrate the existence of multiple bound states and give the dependence of their number on the size and depth of the gap suppression. Extending our
analysis beyond the semiclassical regime, in dimensions larger than one, we observe an infinite number of bound states very close to the gap edge,
even for an arbitrarily small trap. These results deepen our understanding of trapped quasiparticles and may have
important implications for their recombination in disordered superconductors.
\end{abstract}

\maketitle

\input{Introduction}

\input{Model}

\input{Semiclassics}

\input{Beyond-semiclassics}

\input{WaveFunction}

\input{Conclusion}


\appendix

\input{Appendix}


\end{document}

%% file: Introduction.tex
\section{Introduction}

Conventional s-wave superconductors have a gap $\Delta=1.76k_BT_c$, where $T_c$ is the superconducting transition temperature, for quasi-particle excitations around the Fermi level. As a consequence, in thermal equilibrium, the quasiparticle (QP) concentration predicted by microscopic BCS theory exhibits exponential
suppression with decreasing temperature. At milli-Kelvin temperatures, the expected concentration in aluminum would correspond to about two
quasiparticles in a device the size of the Earth~\cite{QP_Poisoning}. This contradicts experimental observations: Various experiments on superconducting devices report a significant excess of the quasiparticle population~\cite{Energy_decay_Qubit/High_QP_pop,High_QP_pop,QP_cavities_1,QP_trapped_vortices}. Such an excess may arise due to a variety of pair-breaking perturbations such as phonons~\cite{phonon_QP_1,phonon_QP_2}, IR photons~\cite{IR-photons,env_radiation}, or cosmic radiation~\cite{ionizing_radiation,Qubit_under_mountain}, some of which are difficult to control.
The excess quasiparticles pose significant challenges for superconducting applications as they reduce qubit lifetimes and coherence times~\cite{Energy_decay_Qubit/High_QP_pop, Catelani1,T1_QUbit,T2_Qubit_QP} and  degrade the quality factor of superconducting resonators~\cite{QP_cavities_1}.  While various mitigation strategies exist~\cite{QP_trapped_vortices,mitigation0,mitigation1,mitigation2,phonon_QP_1,gap-engineering}, further research is needed. In particular, it is crucial to understand the  quasiparticle dynamics and their recombination in realistic devices. 

At low quasiparticle concentrations, recombination is slow as two quasiparticles are required. The process further slows down in the presence of disorder, which leads to small variations in the gap amplitude that may trap quasiparticles~\cite{Bespalov_1}.  To go beyond the model of Ref.~\cite{Bespalov_1}, a more thorough modeling of the quasiparticle dynamics is necessary. Since the energy spectrum plays a key role in quasiparticle dynamics, it must be carefully investigated. Ref.~\cite{Bespalov_1} used the Larkin-Ovchinnikov theory~\cite{Larkin-Ovchinnikov} of optimal fluctuations. According to the work of Larkin and Ovchinnikov, the density of states at an energy $E_b$  below the gap edge is proportional to  the probability to realize a gap suppression sufficiently deep such that the lowest lying state in the suppression possesses energy $E_b$. While their considerations are sufficient to obtain the density of states, the full spectrum of bound states in the gap suppression is necessary in order to address the dynamics. While some prior work exists~\cite{Weinkauf-Zittartz,Flatte-Byers,Gunsenheimer,Chattopadhyay,Andersen,Bespalov2}, a comprehensive analysis of the bound state spectrum as a fonction of the size of the gap variation over different length scales is missing.

Bound states in superconductors are ubiquitous.  In addition to the above mentioned gap variations due to weak potential disorder, they arise also in the presence of magnetic impurities, where they are called Yu-Shiba-Rusinov (YSR) states~\cite{Yu,Shiba,Rusinov} . The spectrum of bound states—including the number, degeneracy, and energy dispersion—depends sensitively on the strength and spatial extent of the perturbation that creates them. While the optimal fluctuations  extend over scales much larger than the superconducting coherence length, magnetic impurities are typically modeled as $\delta$-potentials. 

Here we concentrate on a simple toy model, where the gap is reduced from $\Delta_0$ to $\Delta_0-\delta\Delta$ in a region of radius $R$, to address the bound state spectrum in one, two, and three dimensions for an arbitrary radius $R$. We find that in a gap suppression of size $R>\lambda_F$, where $\lambda_F$ is the Fermi wavelength, semiclassics is sufficient to describe the states sufficiently far from the edge of the superconducting gap. For a trap the size of the superconducting coherence length, the typical scale over which the gap varies, semiclassics yields a large number of bound states in two and three dimensions in contrast to only a few bound states in one dimension.

By contrast, in the limit $R\to0$, the lowest-lying bound state is recovered by assuming a $\delta$-function potential. However, we show that additional bound states closer to the gap edge exist that neither of these descriptions captures. To describe these states, we study the complete energy spectrum of bound states by solving the Bogoliubov–de Gennes equations. We show that in two and three dimensions, an infinite number of bound states corresponding to arbitrarily high angular momentum exist for any finite trap size. These states accumulate very close to the gap edge. Their spatial profile is unusual in that the maximum probability density may be far away from the trap position at a radius $r\sim \ell\lambda_F$, where $\ell$ is the orbital angular momentum quantum number.

The paper is structured as follows. We start by introducing the model in Sec.~\ref{sec-model}. Then, in Sec.~\ref{sec-BS}, we use semiclassics to compute the bound state energy spectrum and determine the number of bound sates as a function of the radius of the gap suppression. In Sec.~\ref{sec-BdG}, we use the full Bogoliubov-de Gennes equations to bridge the previous results with the limit of a $\delta$-impurity and show that additional bound states exist. In Sec.~\ref{sec-wf}, we discuss the wave functions of these additional states, before concluding in Sec.~\ref{sec-conclusion}. Some additional details can be found in the appendices.

%% file: Model.tex
\section{The model}
\label{sec-model}

We consider a conventional superconductor described by the mean-field Bogoliubov-de Gennes Hamiltonian
\begin{equation}
{\cal  H}_{\rm BdG}=\begin{pmatrix}-\frac1{2m}\vec\nabla^2-\mu&\Delta(\vec r)\\ \Delta^*(\vec r)&\frac1{2m}\vec\nabla^2+\mu\end{pmatrix}.
\label{eq-BdG}
\end{equation}
Here $m$ is the effective electron mass, $\mu$ is the chemical potential, and we chose units where $\hbar=1$.
The gap can be chosen real, and we assume a gap profile
\begin{equation}\Delta(\vec r)=\Delta_0-\delta\Delta\,\theta(R-|\vec r|),\label{eq-gap}\end{equation}
where $\theta(x)$ is the Heaviside function and $0<\delta\Delta\ll\Delta_0$. {The exact radial dependence should not matter as the length scale $R_c$ characterizing the spatial extent of the bound states is large compared to the superconducting coherence length $\xi_0=v_F/\Delta_0$, where $v_F$ is the Fermi velocity, over which the gap variation would be smoothened. Namely $R_c\sim\xi=v_F/\sqrt{2\Delta_0\delta\Delta}$ as we will see below.} 

Since we are interested in bound states that form at energies $\Delta_0-\delta\Delta <E<\Delta_0$, we may use an effective Hamiltonian
\begin{equation}
{\cal H}_+=-\frac{(\vec\nabla^2+2m\mu)^2}{8m^2\Delta_0}+\Delta(\vec r)
\label{eq-Heff}
\end{equation}
to describe states in that energy range. In particular, we will search for solutions of the stationary Schrödinger equation
\begin{equation}\left(\frac{(\vec\nabla^2+2m\mu)^2}{8m^2\Delta_0{\delta\Delta}}+\,\theta(R-|\vec r|)\right)\psi(\vec r)={\epsilon}\psi(\vec r),
\label{eq-schroedinger}
\end{equation}
where {$\epsilon=(\Delta_0-E)/\delta\Delta$} is the {dimensionless} binding energy {defined within the range $0<\epsilon<1$}.

%% file: Semiclassics.tex
\section{Semiclassics}
\label{sec-BS}

To obtain the semiclassical result for the bound states, we need the Andreev reflexion coefficient $r_A$ for quasi-particles with energy $\Delta_0-\delta\Delta<E<\Delta_0$ that impinge from a superconductor with gap $\Delta_0-\delta\Delta$ on a superconductor with gap $\Delta_0$. Matching wave functions at the boundary yields (in the Andreev approximation $\Delta_0\ll \mu$)
\begin{equation}\begin{pmatrix}u\\v\end{pmatrix}+r_A\begin{pmatrix}v\\u\end{pmatrix}=\tilde t\begin{pmatrix}1\\a\end{pmatrix},\label{rA}\end{equation}
where $v/u=(E-\sqrt{E^2-(\Delta_0-\delta\Delta)^2})/(\Delta_0-\delta\Delta)$ and $a=(E-i\sqrt{\Delta_0^2-E^2})/\Delta_0$. Here $\tilde t$ is the amplitude of the evanescent part of the wavefunction outside the gap suppression, whose expression will not be needed. This yields the Andreev reflection coefficient $r_A=(a-v/u)/(1-av/u)$. Thus, the Andreev phase $\phi_A={\rm arg}(r_A)$ is given as
\begin{equation}\phi_A=-\arccos\frac{E^2-\Delta_0(\Delta_0-\delta\Delta)}{E\delta\Delta}.\end{equation}
{Using the dimensionless binding energy $\epsilon$ defined above}, we find
\begin{equation}\phi_A
\approx-\arccos(1-2\epsilon),\end{equation}
where we used $\delta\Delta\ll\Delta_0$. Thus, the phase varies between $\phi_A=0$ at $\epsilon=0$ and $\phi_A=-\pi$ at $\epsilon=1$.

\subsection{One-dimensional case}

In the Andreev approximation states close to $\pm k_F$, where  $k_F=\sqrt{2m\mu}$ is the Fermi  momentum, are decoupled and yield two degenerate solutions. Thus, it is sufficient to consider momenta close to $+k_F$. The electron/hole momenta at energy $E=\Delta_0-\delta\Delta\epsilon$ in the gap suppression are given as
\begin{equation}
k_{e/h}\approx k_F\pm\sqrt{\frac{m\Delta_0{\delta\Delta}}{\mu}{\left(1-\epsilon\right)}}.
\end{equation}
Using the Bohr-Sommerfeld quantization condition, we find
\begin{equation}-2\arccos(1-2\epsilon)+4\gamma s\sqrt{1-\epsilon}=2n\pi, \qquad n\in\mathbb{N},\label{sc-1D}\end{equation}
where we introduced the following dimensionless parameters:
\begin{equation}s=k_FR,\qquad \gamma=\sqrt{\Delta_0\delta\Delta/2\mu^2}\ll1.\end{equation}
Note that $\gamma =(k_F\xi)^{-1}$ with the length $\xi$ introduced below Eq.~\eqref{eq-gap}, such that $\gamma s=R/\xi$. 

To count the number of bound states, it is sufficient to study when they enter the trap at energies $\epsilon \ll 1$, so that we can perform a Taylor expansion of \eqref{sc-1D}. 

When $\gamma s\ll1$,  {the first localized state, corresponding to $n=0$, has energy} 
$\epsilon_0=\gamma^2s^2$.
Additional states with $n>0$ appear at the threshold values $s_n=n\pi/(2\gamma)$, that is $R_n=n\pi \xi_0/(2\sqrt{2\delta\Delta/\Delta_0})$. 
At $\gamma s\gg1$, there are thus many bound states with energies
$\epsilon_n=1-\left(n\pi/2\gamma s\right)^2$
far from threshold.

The total number of bound states as a function of trap size  is  given as $N_{\rm 1D}(s)=4\gamma s/\pi$ or $N_{\rm 1D}(R)=2\nu_{\rm 1D}\sqrt{2\Delta_0\delta\Delta}2R$ {with $\nu_{\rm 1D}=1/(v_F\pi)$ the normal state density of states per spin in one dimension.}

\subsection{Higher dimensions}

	\begin{figure}[h!]
		\includegraphics[width=0.6\columnwidth]{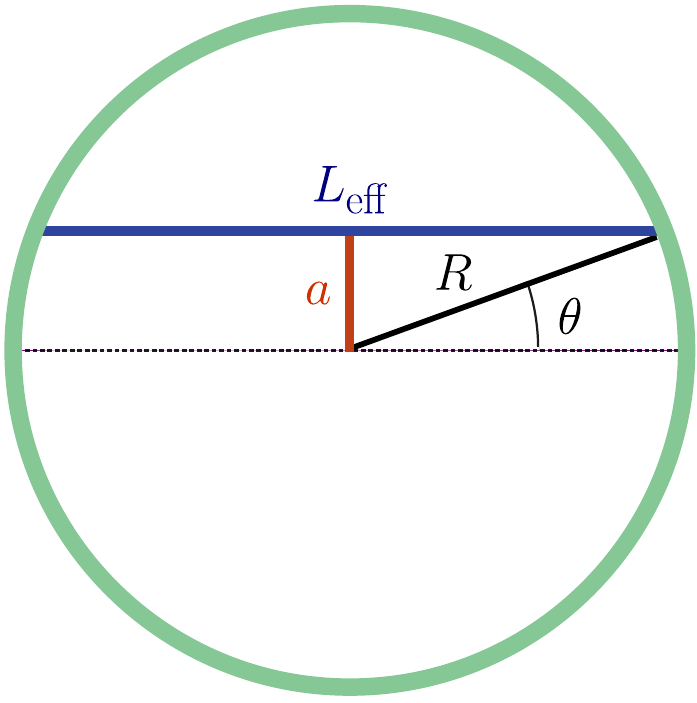}
		\caption{In two dimensions, different semiclassical trajectories are possible. They are characterized by their \lq\lq impact parameter\rq\rq\ $a=R\sin\theta$ (red), which determines the angular momentum $\ell=k_Fa$ of the corresponding bound state. The path length of the trajectory is then given by $2L_{\rm eff}$ with the effective length $L_{\rm eff}=2R\cos\theta$ (blue).\label{fig-2D}}
	\end{figure}
	
We now turn to the two-dimensional case. We may again use the Bohr-Sommerfeld quantization condition. Now there are different path lengths depending on the \lq\lq impact parameter\rq\rq, i.e., the distance $a=R\sin\theta$  between the semiclassical  trajectory and the center of the disk, see Fig.~\ref{fig-2D}. Here $\theta$ is related to the angular momentum $\ell=k_FR\sin\theta\in\mathbb{Z}$. The corresponding path length is $4R\cos\theta=4R\sqrt{1-(\ell/k_FR)^2}$.  Thus, instead of Eq.~\eqref{sc-1D}, we find
\begin{equation}-2\arccos(1-2\epsilon)+4\gamma\sqrt{s^2-\ell^2}\sqrt{1-\epsilon}=2n\pi\label{sc-2D}\end{equation}
with $n\in\mathbb{N}$.

We first note that at $s<1$, a single state with $n,\ell=0$ and energy
$\epsilon_0=\gamma^2s^2$
exists. Additional states appear at the threshold values
\begin{equation}s_{n,\ell}=\sqrt{n^2\pi^2/(4\gamma^2)+\ell^2} .\label{eq-seuils}\end{equation}
Close to threshold, their energy is given as
\begin{eqnarray}
\epsilon_{n,\ell}\approx \gamma^2\delta s_{n,\ell}^2\left(1+\frac{4\gamma^2\ell^2}{n^2\pi^2}(1-\delta_{n,0})\right)
\label{BS-energy}
\end{eqnarray}
with $\delta s_{n,\ell}=s-s_{n,\ell}\ll1/\gamma$. 
Far from threshold, at $\gamma\delta s_{n,\ell}\gg1$,  their energy is given as
\begin{equation}\epsilon_{n,\ell}\approx 1-\left(\frac{n\pi}{2\gamma \sqrt{s^2-\ell^2}}\right)^2.\end{equation}
States $\ell$ and $-\ell$ are degenerate.

Thus, the total number of states as a function of the trap size is given as
\begin{equation}N_{\rm 2D}(s)=2\sum_{n<2\gamma s/\pi}\sqrt{s^2-\left(\frac{\pi}{2\gamma}n\right)^2}=\gamma s^2,\end{equation}
that is
$N_{\rm 2D}(R)=2\nu_{\rm 2D}\sqrt{2\Delta_0\delta\Delta}\,\pi R^2$
with $\nu_{\rm 2D}=m/(2\pi)$ the normal state density of states per spin in two dimensions.

The three-dimensional case has been addressed in Ref.~\cite{Bespalov2}. The energies of the bound states are the same as in two dimensions, but their degeneracy is $2\ell+1$ ($\ell\geq0$). 
Thus, the total number of bound states as a function of the trap size is 
\begin{equation}N_{\rm 3D}(s)=\sum_{n<2\gamma s/\pi}\sum_{\ell<\sqrt{s^2-\left(\frac{\pi}{2\gamma}n\right)^2}}(2\ell+1)=\frac4{3\pi}\gamma s^3,\end{equation}
that is
$N_{\rm 3D}(R)=2\nu_{\rm 3D}\sqrt{2\Delta_0\delta\Delta}\,(4\pi R^3/3)$
with $\nu_{\rm 3D}=mk_F/(2\pi^2)$  the normal state density of states per spin in three dimensions.

The semiclassical result reveals that the higher dimensional case is clearly distinct from the one-dimensional case. As the superconducting gap varies on the scale of the coherence length, the typical size of the trap may be estimated as $\gamma s\sim1$. While in one dimension this yields of order one bound state, in two and three dimensions the number of bound states for that trap size is already large. In particular,
$N_{2D}\sim k_F\xi$ and $N_{3D}\sim (k_F\xi)^2$.

%% file: Beyond-semiclassics.tex
\section{Beyond semiclassics}
\label{sec-BdG}

Semiclassics is expected to well describe the bound state spectrum in sufficiently large traps ($R\gg k_F^{-1}$) and for sufficiently small angular momenta ($k_F R-\ell \ll k_FR$). By contrast, one may note that the results detailed above do not match the bound state energies for a $\delta$-impurity in the limit $\gamma s\ll1$, see Appendix~\ref{sec-delta}. To explore the full parameter range, we thus solve the effective Schrödinger equation \eqref{eq-schroedinger} to obtain the bound state energies. As the potential is piece-wise constant, we search for solutions $\psi_<$ of the effective Schr\"odinger equation  inside the trap and $\psi_>$ outside the trap and then match them at the boundary. 

\subsection{One-dimensional case}

In the one-dimensional case, the effective Schr\"odinger equation reads
\begin{equation}-\frac{(\partial_x^2+2m\mu)^2}{8m^2\Delta_0{\delta \Delta}}\psi(x)=\left[\theta(R-|x|)-{\epsilon}\right]\psi(x).\
\label{BDG_1D}
\end{equation}
The solutions are plane waves, $\psi(x)\propto e^{ikx}$.  Using that $k_F\xi_0\gg1$  and discarding the solutions that diverge at $|x|\to\infty$, the corresponding $k_<$ and $k_>$ at $|x|<R$ and $|x|>R$, respectively,  are given by
\begin{eqnarray}
k_<(\sigma,\sigma')&\approx& \sigma k_F+\sigma'\delta k,\label{eq-kl}\\
 k_>(\sigma)&\approx& \sigma k_F +i\delta p\,{\rm sign}(x),\label{eq-kg}\;\;\;\;\;
\end{eqnarray}
with $\delta k=k_F\gamma\sqrt{1-\epsilon}$, $\delta p=k_F\gamma\sqrt{\epsilon}$, and $\sigma,\sigma'=\pm1$.  

The semiclassical results can be recovered by assuming that  solutions with momenta close to $\pm k_F$ decouple. The wave function matching without this assumption is straightforward though lengthy and yields the following equation for the energies : 
\begin{eqnarray}
&&4\sqrt{\epsilon( 1 - \epsilon )} \left[2 -  \gamma^2(1 - 2 \epsilon)\right] \cos \left( 2\gamma s \sqrt{1 - \epsilon} \right)\\
&&-\left[4(1-2\epsilon) + \gamma^2\left(1-2(1-2\epsilon)^2\right) \right] \sin \left( 2\gamma s\sqrt{1 - \epsilon} \right)\nonumber \\
&=&\pm \gamma \sqrt{1 - \epsilon} \left( 4 \gamma\sqrt{\epsilon} \cos(2 s) - (4 -\gamma^2) \sin(2 s) \right). \nonumber
\label{1D_BdG_full}
\end{eqnarray}

At small trap size $\gamma s\ll1$, we find two states with energies close to the gap edge,
\begin{eqnarray}
\epsilon_\pm=\gamma^2\left(s\mp \frac12\sin(2 s) \right)^2.
\end{eqnarray}
For $s\gg1$, the energies approach the semiclassical result of a doubly-degenerate bound state with energy $\epsilon_0=\gamma^2s^2$. By contrast, for $s\ll1$,  we find two non-degenerate states with energies $\epsilon_0\approx 4\gamma^2s^2$ (corresponding to a $\delta$-potential with weight $U_0=\delta\Delta2R$, see Appendix ~\ref{sec-delta}) and $\epsilon_0'\approx 4\gamma^2s^6/9\ll\epsilon_0$.  Additional states enter at
\begin{eqnarray}
 \sin \left( 2\gamma s\right)=\pm \gamma \sin(2 s),
\end{eqnarray}
so we recover the semiclassical thresholds $s_n$ with a splitting of order 1.	

Thus, in one dimension, semiclassics is sufficient to find the number of bound states. Only the lifting of degeneracy of the bound state in very small traps, $s\ll1$, is beyond semiclassics.

					
\subsection{Higher dimensions}

In the two-dimensional case, we may use the rotational symmetry of the gap suppression to write the wave function in the form $\psi(\vec r)=\psi_r(r)\psi_\theta(\theta)$. Namely,
using $\vec\nabla^2=\partial_r^2+r^{-1}\partial_r+r^{-2}\partial_\theta^2$, we find $\psi_\theta(\theta)=\exp[i\ell\theta]$ with $\ell\in\mathbb{Z}$ and 
\begin{equation}-\frac{(\partial_r^2+\frac1r{\partial_r}-\frac{\ell^2}{r^2}+2m\mu)^2}{8m^2\Delta_0 {\delta \Delta }}\psi_r(r)=\left[\theta(R-r)-{\epsilon}\right]\psi_r(r).\label{eq-2d}\end{equation}
States with $\ell$ and $-\ell$ are degenerate. Thus we concentrate on $\ell\geq0$ below. 

As before, we may search for solutions $\psi_<(r)$ at $r<R$ and $\psi_>(r)$ at $r>R$ and then match them at $r=R$. 
Note that the equation 
\begin{equation}\left(\partial_r^2+\frac1r{\partial_r}-\frac{\ell^2}{r^2}\right)\phi(r)=-k^2\phi(r)\end{equation}
has for solution the Bessel functions $Z_\ell(kr)$, where $Z=J,Y$. These Bessel functions solve Eq.~\eqref{eq-2d} with 
wavevectors inside and outside the trap that have the same form as in the one-dimensional case, Eqs.~\eqref{eq-kl} and \eqref{eq-kg}. It is sufficient to keep the solutions with $\Re[k]>0$, namely
\begin{eqnarray}
\!\!\!\!{k_{<}^\pm}=k_F\left(1\pm\gamma\sqrt{1-\epsilon}\right),\quad
{k_{>}^\pm}=k_F\left(1\pm i \gamma\sqrt{\epsilon}\right).\;\;\;\;
\end{eqnarray}
As  the Bessel functions $Y_\ell(kr)$ diverge at $r\to0$, the solution inside the gap suppression takes the form
\begin{equation}\psi_<(r)=A_+ J_\ell(k_+r)+A_- J_\ell(k_-r),\label{eq-less}\end{equation}
where we introduced the shortened notation $k_\pm = k_{<}^\pm $
Outside  the gap suppression, it is more convenient to use the modified Bessel functions $K_\ell(\pm i kr)$, where $\Re[\pm ik]>0$ to ensure convergence. Defining $p_\pm= \mp ik_{>}^\pm$, the solution outside the gap suppression may thus be written as
\begin{equation}\psi_>(r)=B_+ K_\ell(p_+r) +B_- K_\ell(p_-r). \label{eq-psi_out}\end{equation}
As we have a fourth-order differential equation, we need to impose continuity of the wave function and its first three derivatives. After a lengthy calculation, we obtain
\begin{widetext}
\begin{eqnarray}
&&(k_+^2+p_-^2)(k_-^2+p_+^2)\left[k_-J_{\ell+1}^-K_{\ell}^--p_-J_{\ell}^-K_{\ell+1}^-\right]\left[k_+J_{\ell+1}^+K_{\ell}^+-p_+J_{\ell}^+K_{\ell+1}^+\right]\label{eq-sol}\\
&=&(k_+^2+p_+^2)(k_-^2+p_-^2)\left[k_-J_{\ell+1}^-K_{\ell}^+-p_+J_{\ell}^-K_{\ell+1}^+\right]\left[k_+J_{\ell+1}^+K_{\ell}^--p_-J_{\ell}^+K_{\ell+1}^-\right],\nonumber
\end{eqnarray}
\end{widetext}
where we denoted  $Z_l^\pm =Z_l(q_\pm R)$ for $q=k,p$.
Expanding Eq.~\eqref{eq-sol} in $\gamma\ll s^{-1}$, we obtain 
\begin{eqnarray}
\epsilon_\ell =\frac{\pi^2 \gamma^2s^2}4 
\left[s(J_\ell^2(s)+J_{\ell+1}^2(s))- 2 \ell J_\ell(s) J_{\ell+1}(s)\right]^2\!\!,\;\;\;\;\;\label{res-simple}
\end{eqnarray}
using that $\epsilon\ll1$ in this regime.

Thus, we find a bound state for any value of $\ell$ at arbitrarily small $s$, in marked contrast to the semiclassical result.
In particular, for $s\ll1$, expanding the Bessel functions at small argument, $J_\ell(z)\approx(z/2)^\ell/\ell!$, yields
\begin{eqnarray}
\epsilon_\ell &\approx&\frac{\pi^2 s^{4(\ell+1)}\gamma^2}{2^{4\ell+2}(\ell !(\ell+1)!)^2}
=\left(\frac{(s/2)^{2\ell}}{\ell !(\ell+1)!}\right)^2\epsilon_0.
\label{n=0_equation}
\end{eqnarray}
For the lowest-lying state, $\ell=0$, we have $\epsilon_0=\pi^2\gamma^2s^4/4$, which  matches the result for a $\delta$-potential with weight $U_0=\delta\Delta \pi R^2$, see Appendix~\ref{sec-delta}. The state  $\ell=1$ has energy $\epsilon_1=\epsilon_0s^4/64$, i.e., the level spacing between the two lowest states is $\epsilon_1-\epsilon_0\approx\epsilon_0$. The energies of states with $\ell\gg1$ are given as
\begin{equation}\epsilon_\ell\approx\epsilon_0\exp\left[-4\ell\ln\frac {2\ell}s\right],\end{equation}
such that the states are exponentially close to the gap edge. 

The semiclassical result is recovered only at $s\gg\ell$. In that case, the Bessel functions may be approximated as
$J_\ell(s)\approx\sqrt{2/{\pi s}}\cos(s-\pi(2\ell+1)/4)$,
yielding
\begin{eqnarray}
\epsilon_\ell =\gamma^2s^2\left[1-2\frac\ell s\sin(2s-\frac\pi2(2\ell+1))\right]\approx \gamma^2s^2.
\end{eqnarray}
As in the one-dimensional case, on top of the semiclassical result, there are oscillations with small amplitude $\delta\epsilon/\epsilon\sim s^{-1}$.

\begin{figure}[h!]
\includegraphics[width=1\columnwidth]{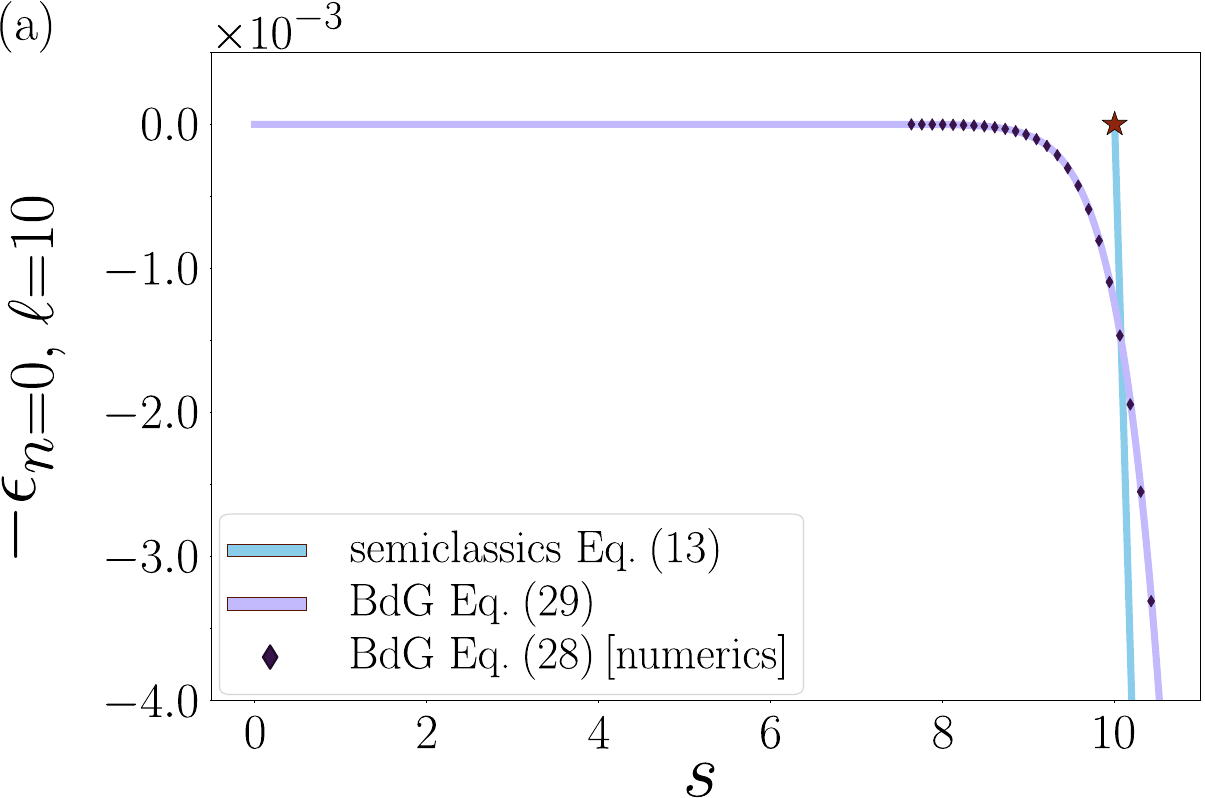}
\includegraphics[width=1\columnwidth]{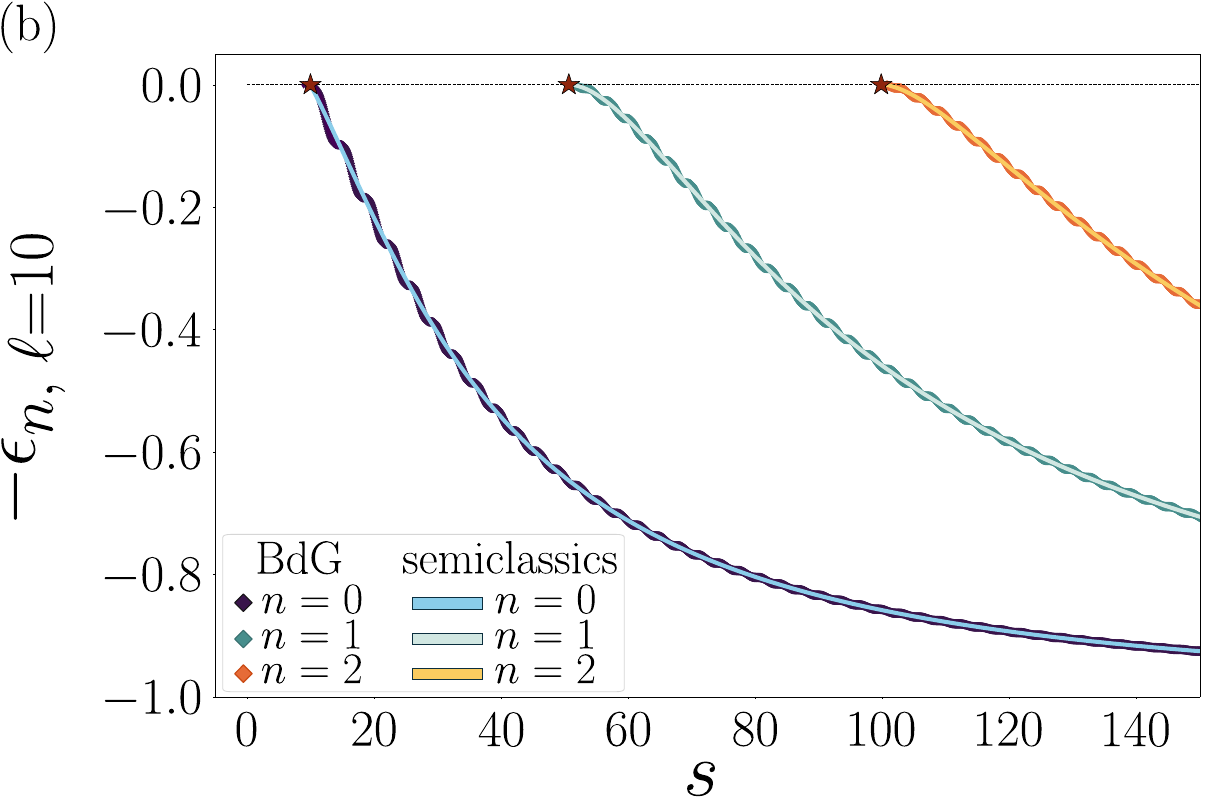}
    \caption{Bound state energies  for $\ell=10$ {and $\gamma^2=10^{-3}$} as a function of trap size $s$ in a two-dimensional trap. (a) Energy of the lowest-lying bound state at small trap sizes. Analytical results obtained with semiclassics (Eq.~\eqref{BS-energy}) and with the full Bogoliubov-de Gennes equation (Eq.~\eqref{res-simple}), as well as numerical results using Eq.~\eqref{eq-sol} are shown.  The semiclassical threshold value is indicated by a star.  While the numerics show how the bound state detaches from the gap edge around the semiclassical threshold, the numerical accuracy is not sufficient to resolve the bound state energy at much smaller trap sizes. 
    (b) Energy spectrum of the first few bound states showing good agreement between the  semiclassical result, Eq.~\eqref{sc-2D}, and  the BdG result, Eq.~\eqref{eq-sol}, at larger scales. 
    \label{fig-spectrum2D}}
\end{figure}

The numerical solution of Eq.~\eqref{eq-sol} for $\ell=10$ is shown in Fig.~\ref{fig-spectrum2D}. While the energies close to the gap edge at very small trap size are beyond the numerical accuracy, the persistence of the bound state for trap sizes below the semiclassical threshold is clearly seen in Fig.~\ref{fig-spectrum2D}(a). The state detaches from the gap edge around the semiclassical threshold value. The agreement with the semiclassical result at larger trap sizes is shown in  Fig.~\ref{fig-spectrum2D}(b).

\begin{figure}
	\includegraphics[width=1\columnwidth]{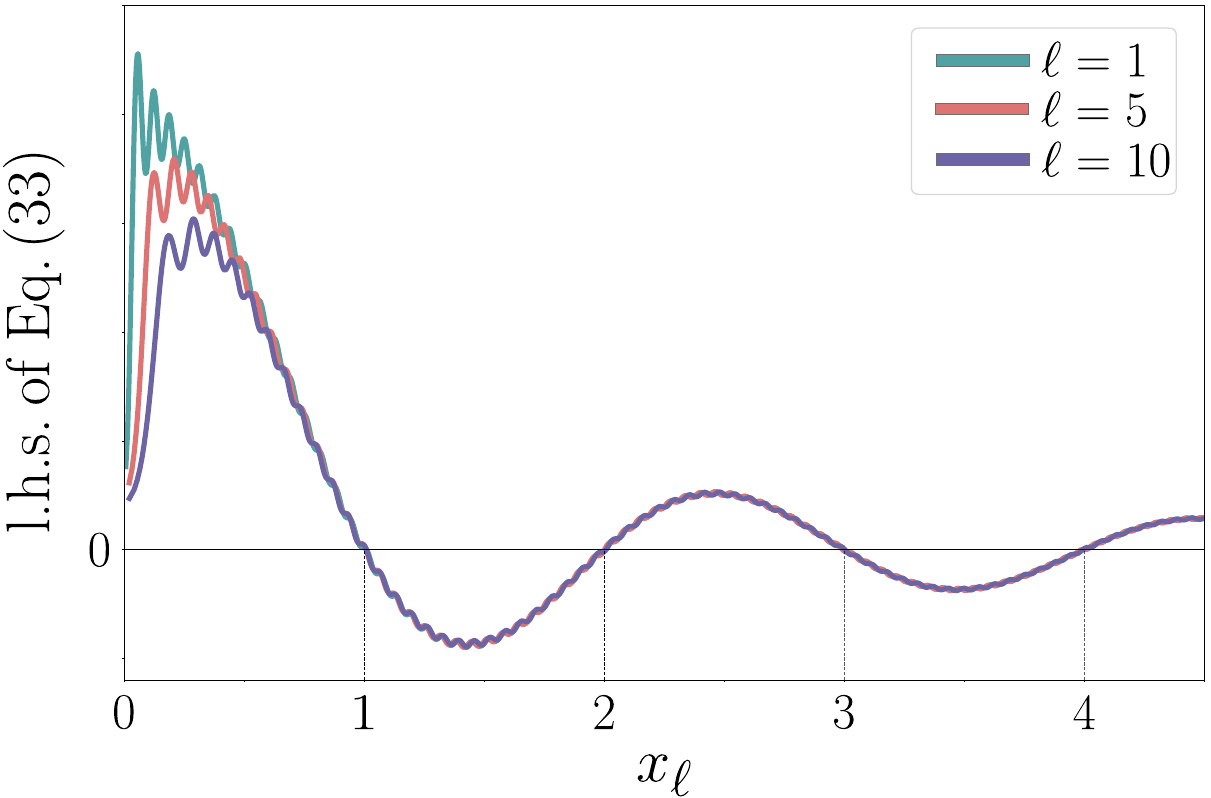}
	\caption{Left-hand side of Eq.~\eqref{s-threshold} for $\ell=1,5,10$ {and  $\gamma^2=10^{-3}$} as a function of $x_\ell=2\gamma/\pi\sqrt{s^2-\ell^2}$. Thresholds are given by zero crossings. All the curves cross zero very close to the semiclassical thresholds $s_{n,\ell}$ corresponding to $x_\ell=n$ with $n>0$. \label{fig-seuils}}
\end{figure}

Fig.~\ref{fig-spectrum2D}(b) also shows additional states with the same orbital momentum that appear at larger trap sizes. The threshold values $s_c$ can be found by solving Eq.~\eqref{eq-sol} at $\epsilon=0$. Namely, the equation simplifies to
\begin{eqnarray}
(1-\gamma)J_\ell((1+\gamma){s_c})J_{\ell+1}((1-\gamma){s_c})&&\label{s-threshold}\\
-(1+\gamma)J_\ell((1-\gamma){s_c})J_{\ell+1}((1+\gamma){s_c})&=&0.\nonumber
\end{eqnarray}
For ${s_c}\gg\ell$, we may approximate the Bessel functions by their asymptotic expressions yielding
\begin{equation}
(-1)^\ell\gamma\cos(2{s_c})+\sin(2\gamma {s_c})=0.
\end{equation}
Thus, for $\gamma\ll1$, we recover the semiclassical threshold values for $\ell=0$, namely
$s_{n,0}\approx n\pi/(2\gamma)$.
The $\ell$-dependence of the threshold values is beyond this approximation. Fig.~\ref{fig-seuils} shows the left-hand side of Eq.~\eqref{s-threshold} plotted as a function of $x_\ell=(2\gamma/\pi)\sqrt{s^2-\ell^2}$ such that the semiclassical thresholds $s_{n,\ell}$ correspond to $x_\ell=n$.
As one can see, for $n>0$ and $\gamma\ll1$, the zeros of the left-hand side of Eq.~\eqref{s-threshold}  are very close to the semiclassical thresholds for all values of $\ell$. 

The appearance of an infinite number of bound states at arbitrarily small  trap size was also noted in Ref.~\cite{Bespalov2} for the three-dimensional case. In that case their energies at $s\ll1$ were found to be given as
\begin{equation}\epsilon_\ell=4\gamma^2\frac{s^{4\ell+6}}{[(2\ell+1)!!(2\ell+3)!!]^2}\label{3D_BdG_E}\end{equation}
(see Eq. (C15) in App.~C of Ref.~\cite{Bespalov2}).  In particular, the energy of the state $\ell=0$ is
$\epsilon_0=4\gamma^2s^{6}/{9}$, corresponding to a $\delta$-potential with $U_0=\delta\Delta 4\pi R^3/3$, see Appendix~\ref{sec-delta}. As in the two-dimensional case, states with $\ell\gg1$ are exponentially close to the gap edge.

\
\\

Thus, the differences between the semiclassical result and the solution of the full Bogoliubov-de Gennes equation are more pronounced in higher dimensions than in the one-dimensional case. Namely, the absence of a threshold for the first bound state for each value of $\ell$ yields an infinite number of bound states for arbitrary  trap sizes. By contrast, semiclassics yields a finite number of bound states at arbitrary trap size. In the next section, we are going to further characterize the bound states close to the gap edge that are beyond semiclassics by studying their wavefunctions.

%% file: WaveFunction.tex
\vspace{1cm}
\section{Wave functions}
\label{sec-wf}

The same matching procedure used to obtain the bound state energies allows one also to obtain the bound state wave functions. Details are given in Appendix~\ref{app-wf}.

The decay of the wave function outside the trap is determined by the imaginary part of the wavevector $\Im[k_>]=k_F\gamma\sqrt{\epsilon}$. Namely, the decay length is $R_c=\xi/\sqrt{\epsilon}$. Thus, the closer the energy to the gap edge, the slower the decay. In particular, the states beyond semiclassics that are exponentially close to the gap edge decay on an exponentially long length scale. At smaller scales, $r\ll R_c$, we may neglect the imaginary part of  the wavevector. In that regime, the wave function outside the trap, Eq.~\eqref{eq-psi_out}, takes the form
\begin{equation}\psi_>(r)=B_+ K_\ell(-ik_Fr) +B_- K_\ell(ik_Fr).\end{equation}
For $\gamma s\ll1$ such that  $\epsilon\ll1$ and $\delta p\ll\delta k$, the coefficients $B_\pm$ may be approximated as  $B_\pm\approx(\pm1)^{\ell+1}C_B(\ell)$ (see Appendix~\ref{app-wf}), where the normalization constant $C_B(\ell)$ is not necessary to extract the shape of the wave function outside the gap. This yields
\begin{equation}\psi_>(r)=C_B(\ell)\pi i^{\ell+1} J_\ell(k_Fr)\label{eq-wff}\end{equation}
at $r\ll R_c$.

Eq.~\eqref{eq-wff} yields the shape of the wave function at length scales $r\ll R_c$. The Bessel function $J_\ell(x)$ has a global maximum at $x\sim\ell$. Thus, when $\ell >s$, the states form a \lq\lq ring\rq\rq\ with radius $r\sim \ell k_F^{-1}$ around the trap. This behavior is markedly different from the radial dependence of the semiclassical states where the maximal weight of the wave function is within or close to the border of the gap suppression. 

\begin{figure}[H]
		\includegraphics[width=1\columnwidth]{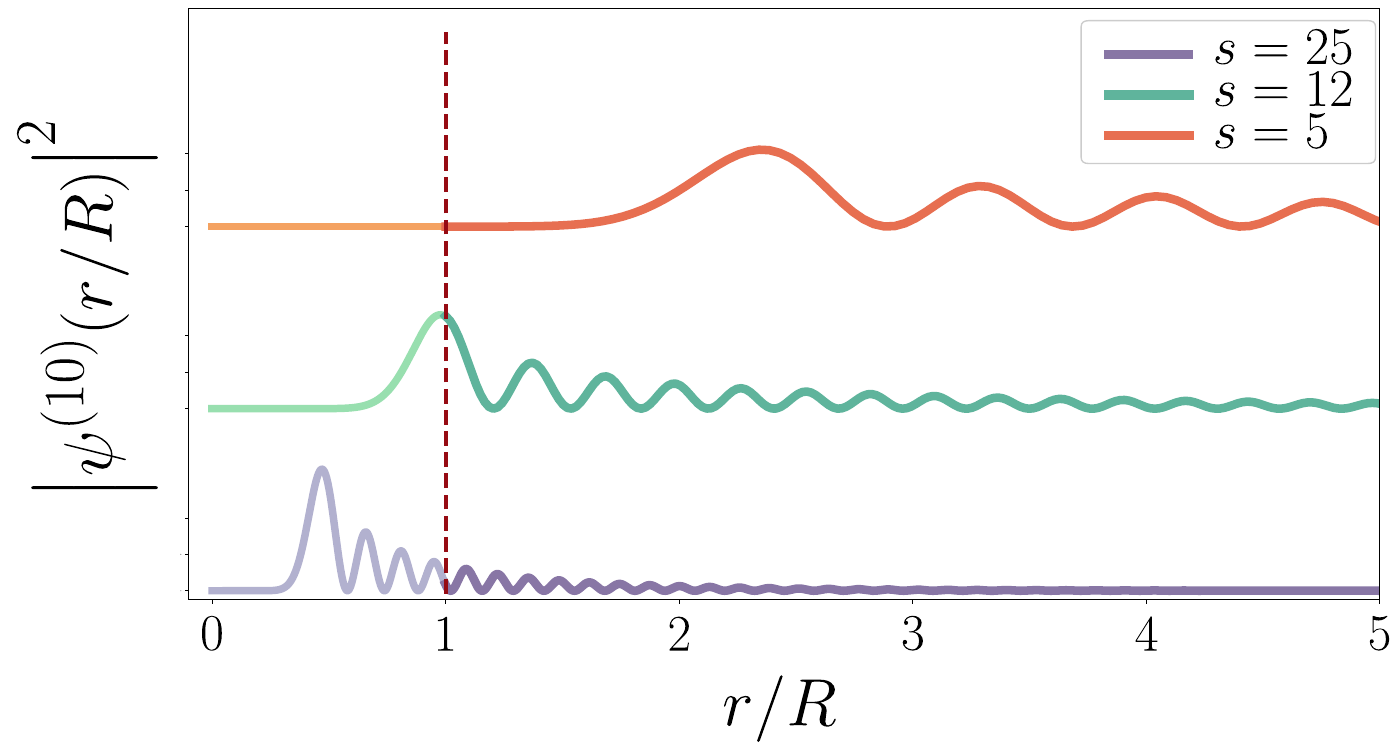}
		\caption{Wave functions for $\ell=10$, {$\gamma^2=10^{-3}$}, and different trap sizes. We show the probability density $|\psi^{(10)}(r/R)|^2$ in arbitrary units for different trap sizes. 
		As $s$ decreases, the maximum of $|\psi^{(10)}(r/R)|^2$ moves outward. For states that are beyond the semiclassical approximation, $s\ll s_{n\ell}$, it lies far outside the trap (red).
		\label{fig-wave}}
	\end{figure}

This behavior is illustrated in Fig.~\ref{fig-wave} for the case $\ell=10$ such that the semiclassical threshold is $s_{0,10}=10$. For $s\gg s_{0,10}$ the maximal weight lies well within the trap. Such states are well described by the semiclassical analysis. For $s\sim s_{0,10}$ the maximal weight is close to the trap edge, whereas for $s\ll s_{0,10}$ it lies far outside the trap. These states are beyond the semiclassical analysis, but captured by the full Bogoliubov-de Gennes equation.

%% file: Conclusion.tex
\section{Conclusion}
\label{sec-conclusion}

We studied a simple toy model for a superconducting gap suppression that may trap quasiparticles. The full Bogoliubov-de Gennes  calculation allows one to recover the semiclassical results for extended gap suppressions as well as the results for a $\delta$-potential in the case of sufficiently small gap suppressions. However, in two and three dimensions, additional states exist close to the gap edge at arbitrary size of the gap suppression that are not captured by either of these approaches, as previously noted in Ref.~\cite{Bespalov2} for the three-dimensional case. We revealed the unusual spatial structure of these additional states that are mainly localized far away from the gap suppression.  

While our model is an oversimplification for describing the shape of the gap variation in a superconductor due to disorder, we expect that the precise shape is not crucial due to the large spatial extension of the bound state wave functions. More realistic models would need to include short-scale disorder and deviations from the rotational symmetry of the gap variation, which are beyond the scope of this work. As typical gap variations in weakly disordered superconductors are on a scale much larger than the superconducting coherence length, the bound state spectrum in this case is quasi-continuous. Even for a minimal trap size of the order of the coherence length, the number of bound states is large in two and three dimensions. This observation may be used as a starting point for the investigation of the quasiparticle dynamics and challenges the claim that quasiparticle traps may act as two-level systems~\cite{balatsky-TLS,ioffe-TLS}. 

\section*{Acknowledgments}

We acknowledge funding
from the Plan France 2030 through the project ANR-22-
PETQ-0006.

%% file: Appendix.tex
\section{$\delta$-impurity}
\label{sec-delta}

In the limit $R\to0$, we may compare our results with the energy of the single (only spin-degenerate) bound state generated by a $\delta$-potential. In that case, one expects
$\epsilon=2 \Delta_0(\pi\nu)^2U_0^2/\delta\Delta$, where the impurity strength $U_0$ is given by the the depth of the gap suppression $\delta\Delta$ integrated over its size,
\begin{eqnarray}
U_0=\delta\Delta\times\begin{cases}2R&{\rm 1D},\\\pi R^2&{\rm 2D},\\4\pi R^3/3&{\rm 3D}.\end{cases}
\end{eqnarray}
The densities of states in different dimensions are given in Sec.~\ref{sec-BS}.
This yields the energy
\begin{eqnarray}
\epsilon=\gamma^2\times\begin{cases}4s^2&{\rm 1D},\\{\pi^2}s^4/4&{\rm 2D},\\4s^6/9&{\rm 3D}.\end{cases}
\end{eqnarray}
These results are recovered by the full Bogoliubov-de Gennes calculation, but not by semiclassics which predicts that the lowest energy bound state at small trap sizes has energy $\epsilon_0=\gamma^2 s^2$ independent of dimensionality.

\section{Wave functions in two dimensions}
\label{app-wf}

In Sec.~\ref{sec-wf}, we study the wave functions for the two-dimensional case. The coefficients $A_\pm$ and $B_\pm$ introduced in Eqs.~(\ref{eq-less},\ref{eq-psi_out}) are provided here up to an overall normalization constant. Namely,
\begin{eqnarray}
A_+&=&-{\cal N}_A(k_-^2+p_-^2)\left[k_-J_{\ell+1}^-K_{\ell}^+-p_+J_{\ell}^-K_{\ell+1}^+\right]\!,\;\;\\
A_-&=&{\cal N}_A(k_+^2+p_-^2)\left[k_+J_{\ell+1}^+K_{\ell}^+-p_+J_{\ell}^+K_{\ell+1}^+\right]\!,\\
B_+&=&-{\cal N}_B(k_-^2+p_-^2)\left[k_+J_{\ell+1}^+K_{\ell}^--p_-J_{\ell}^+K_{\ell+1}^-\right]\!,\;\;\\
B_-&=&{\cal N}_B(k_-^2+p_+^2)\left[k_+J_{\ell+1}^+K_{\ell}^+-p_+J_{\ell}^+K_{\ell+1}^+\right]
\end{eqnarray}
with
\begin{equation}
\frac{{\cal N}_A}{{\cal N}_B}=\frac{(k_-^2+p_+^2)\left[p_-K_{\ell+1}^-K_{\ell}^+-p_+K_{\ell}^-K_{\ell+1}^+\right]}{(k_+^2-k_-^2)\left[k_-J_{\ell+1}^-K_{\ell}^+-p_+J_{\ell}^-K_{\ell+1}^+\right]}.
\end{equation}
It will be of particular interest to study the behavior of the wavefunctions outside the gap suppression. Concentrating on states close to the gap edge, $\epsilon\ll 1$, we may approximate
\begin{widetext}
\begin{eqnarray}
B_\pm&\approx&\pm2{\cal N}_Bk_F^2\delta k\left[J_{\ell+1}(s(1+\gamma))K_{\ell}(\pm is)\mp iJ_{\ell}(s(1+\gamma))K_{\ell+1}(\pm is)\right]
\end{eqnarray}
\end{widetext}
At $s\gamma \ll1$, the expression further simplifies to
\begin{eqnarray}
B_\pm&\approx&\mp (\mp i)^{\ell}\frac{k_F^2\delta k}s
\end{eqnarray}
ou $B_\pm\approx(\pm1)^{\ell+1}C_B(\ell)$, where the normalization constant $C_B(\ell)$ is not necessary to extract the shape of the wavefunction outside the gap.

%% file: main.bbl
\begin{thebibliography}{31}

\bibitem{QP_Poisoning}
J.~Aumentado, G.~Catelani, and K.~Serniak,
\emph{Physics Today} \textbf{76}, 34 (2023).

\bibitem{Energy_decay_Qubit/High_QP_pop}
J.~M.~Martinis, M.~Ansmann, and J.~Aumentado,
\emph{Phys. Rev. Lett.} \textbf{103}, 097002 (2009).

\bibitem{High_QP_pop}
P.~J.~de~Visser, J.~J.~A.~Baselmans, P.~Diener, S.~J.~C.~Yates,
A.~Endo, and T.~M.~Klapwijk,
\emph{Phys. Rev. Lett.} \textbf{106}, 167004 (2011).

\bibitem{QP_cavities_1}
P.~J.~de~Visser, D.~J.~Goldie, P.~Diener, S.~Withington,
J.~J.~A.~Baselmans, and T.~M.~Klapwijk,
\emph{Phys. Rev. Lett.} \textbf{112}, 047004 (2014).

\bibitem{QP_trapped_vortices}
C.~Wang, Y.~Y.~Gao, I.~M.~Pop, U.~Vool, C.~Axline, T.~Brecht,
R.~W.~Heeres, L.~Frunzio, M.~H.~Devoret, G.~Catelani,
L.~I.~Glazman, and R.~J.~Schoelkopf,
\emph{Nat. Commun.} \textbf{5}, 5836 (2014).

\bibitem{phonon_QP_1}
A.~Bargerbos, L.~J.~Splitthoff, M.~Pita-Vidal, J.~J.~Wesdorp,
Y.~Liu, P.~Krogstrup, L.~P.~Kouwenhoven,
C.~K.~Andersen, and L.~Gr{\"u}nhaupt,
\emph{Phys. Rev. Appl.} \textbf{19}, 024014 (2023).

\bibitem{phonon_QP_2}
E.~Yelton, C.~P.~Larson, V.~Iaia, K.~Dodge, G.~La~Magna,
P.~G.~Baity, I.~V.~Pechenezhskiy, R.~McDermott,
N.~A.~Kurinsky, G.~Catelani, and B.~L.~T.~Plourde,
\emph{Phys. Rev. B} \textbf{110}, 024519 (2024).

\bibitem{IR-photons}
R.~Barends, J.~Wenner, M.~Lenander, Y.~Chen, R.~C.~Bialczak,
J.~Kelly, E.~Lucero, P.~O’Malley, M.~Mariantoni, D.~Sank,
H.~Wang, T.~C.~White, Y.~Yin, J.~Zhao, A.~N.~Cleland,
J.~M.~Martinis, and J.~J.~A.~Baselmans,
\emph{Appl. Phys. Lett.} \textbf{99}, 113507 (2011).

\bibitem{env_radiation}
R.~T.~Gordon, C.~E.~Murray, C.~Kurter, M.~Sandberg, S.~A.~Hall,
K.~Balakrishnan, R.~Shelby, B.~Wacaser, A.~A.~Stabile,
J.~W.~Sleight, M.~Brink, M.~B.~Rothwell, K.~P.~Rodbell,
O.~Dial, and M.~Steffen,
\emph{Appl. Phys. Lett.} \textbf{120}, 074002 (2022).

\bibitem{ionizing_radiation}
A.~P.~Veps{\"a}l{\"a}inen, A.~H.~Karamlou, J.~L.~Orrell,
A.~S.~Dogra, B.~Loer, F.~Vasconcelos, D.~K.~Kim,
A.~J.~Melville, B.~M.~Niedzielski, J.~L.~Yoder,
S.~Gustavsson, J.~A.~Formaggio, B.~A.~VanDevender,
and W.~D.~Oliver,
\emph{Nature} \textbf{584}, 551 (2020).

\bibitem{Qubit_under_mountain}
L.~Cardani, F.~Valenti, N.~Casali, G.~Catelani, T.~Charpentier,
M.~Clemenza, I.~Colantoni, A.~Cruciani, G.~D’Imperio,
L.~Gironi, L.~Gr{\"u}nhaupt, D.~Gusenkova, F.~Henriques,
M.~Lagoin, M.~Martinez, G.~Pettinari, C.~Rusconi, O.~Sander,
C.~Tomei, A.~V.~Ustinov, M.~Weber, W.~Wernsdorfer,
M.~Vignati, S.~Pirro, and I.~M.~Pop,
\emph{Nat. Commun.} \textbf{12}, 2733 (2021).

\bibitem{Catelani1}
G.~Catelani, R.~J.~Schoelkopf, M.~H.~Devoret, and L.~I.~Glazman,
\emph{Phys. Rev. B} \textbf{84}, 064517 (2011).

\bibitem{T1_QUbit}
M.~Lenander, H.~Wang, R.~C.~Bialczak, E.~Lucero, M.~Mariantoni,
M.~Neeley, A.~D.~O’Connell, D.~Sank, M.~Weides, J.~Wenner,
T.~Yamamoto, Y.~Yin, J.~Zhao, A.~N.~Cleland,
and J.~M.~Martinis,
\emph{Phys. Rev. B} \textbf{84}, 024501 (2011).

\bibitem{T2_Qubit_QP}
G.~Catelani, S.~E.~Nigg, S.~M.~Girvin, R.~J.~Schoelkopf,
and L.~I.~Glazman,
\emph{Phys. Rev. B} \textbf{86}, 184514 (2012).

\bibitem{mitigation0}
R.-P.~Riwar, A.~Hosseinkhani, L.~D.~Burkhart, Y.~Y.~Gao,
R.~J.~Schoelkopf, L.~I.~Glazman, and G.~Catelani,
\emph{Phys. Rev. B} \textbf{94}, 104516 (2016).

\bibitem{mitigation1}
I.~M.~Pop, K.~Geerlings, G.~Catelani, R.~J.~Schoelkopf,
L.~I.~Glazman, and M.~H.~Devoret,
\emph{Nature} \textbf{508}, 369 (2014).

\bibitem{mitigation2}
V.~Iaia, J.~Ku, A.~Ballard, C.~P.~Larson, E.~Yelton,
C.~H.~Liu, S.~Patel, R.~McDermott, and B.~L.~T.~Plourde,
\emph{Nat. Commun.} \textbf{13}, 6425 (2022).

\bibitem{gap-engineering}
M.~McEwen, K.~C.~Miao, J.~Atalaya, A.~Bilmes, A.~Crook,
J.~Bovaird, J.~M.~Kreikebaum, N.~Zobrist, E.~Jeffrey,
B.~Ying, A.~Bengtsson, H.-S.~Chang, A.~Dunsworth, J.~Kelly,
Y.~Zhang, E.~Forati, R.~Acharya, J.~Iveland, W.~Liu, S.~Kim,
B.~Burkett, A.~Megrant, Y.~Chen, C.~Neill, D.~Sank,
M.~Devoret, and A.~Opremcak,
\emph{Phys. Rev. Lett.} \textbf{133}, 240601 (2024).

\bibitem{Bespalov_1}
A.~Bespalov, M.~Houzet, J.~S.~Meyer, and Y.~V.~Nazarov,
\emph{Phys. Rev. Lett.} \textbf{117}, 117002 (2016).

\bibitem{Larkin-Ovchinnikov}
A.~I.~Larkin and Y.~N.~Ovchinnikov,
\emph{Sov. Phys. JETP} \textbf{34}, 1144 (1972).

\bibitem{Weinkauf-Zittartz}
A.~Weinkauf and J.~Zittartz,
\emph{Z. Phys. B} \textbf{21}, 135 (1975).

\bibitem{Flatte-Byers}
M.~E.~Flatt\'e and J.~M.~Byers,
\emph{Phys. Rev. B} \textbf{56}, 11213 (1997).

\bibitem{Gunsenheimer}
U.~Gunsenheimer and A.~H.~Hahn,
\emph{Physica B} \textbf{218}, 141 (1996).

\bibitem{Chattopadhyay}
A.~K.~Chattopadhyay, R.~A.~Klemm, and D.~Sa,
\emph{J. Phys.: Condens. Matter} \textbf{14}, L577 (2002).

\bibitem{Andersen}
B.~M.~Andersen, A.~Melikyan, T.~S.~Nunner,
and P.~J.~Hirschfeld,
\emph{Phys. Rev. Lett.} \textbf{96}, 097004 (2006).

\bibitem{Bespalov2}
A.~Bespalov, M.~Houzet, J.~S.~Meyer, and Y.~V.~Nazarov,
\emph{Phys. Rev. B} \textbf{93}, 104521 (2016).

\bibitem{Yu}
L.~Yu,
\emph{Acta Phys. Sin.} \textbf{21}, 75 (1965).

\bibitem{Shiba}
H.~Shiba,
\emph{Prog. Theor. Phys.} \textbf{40}, 435 (1968).

\bibitem{Rusinov}
A.~I.~Rusinov,
\emph{JETP Lett.} \textbf{9}, 146 (1969).

\bibitem{balatsky-TLS}
A.~V.~Shytov, I.~Vekhter, I.~A.~Gruzberg,
and A.~V.~Balatsky,
\emph{Phys. Rev. Lett.} \textbf{90}, 147002 (2003).

\bibitem{ioffe-TLS}
S.~E.~de~Graaf, L.~Faoro, L.~B.~Ioffe, S.~Mahashabde,
J.~J.~Burnett, T.~Lindström, S.~E.~Kubatkin,
A.~V.~Danilov, and A.~Y.~Tzalenchuk,
\emph{Sci. Adv.} \textbf{6}, eabc5055 (2020).

\end{thebibliography}
